%% file: hcha.tex
\documentclass[12pt,a4paper,dvips]{article}

\usepackage{a4p}
\usepackage{cite,mcite}
\usepackage{graphicx}
\usepackage{epsfig}   
\usepackage{physics}
\usepackage{l3_title,ifthen}

\date{August 27, 2003}

\journalname{Phys. Lett. B}
\preprint{2003-054}

\newlength{\capindent}
\newlength{\capwidth}
\newlength{\figwidth}
\def\tntn{\mathrm{\tau^+\nu_\tau\tau^-\nbar_\tau}}
\def\cstn{\mathrm{c\bar{s} \tau^-\nbar_\tau}}
\def\cscs{\mathrm{c\bar{s} \bar{c}s}}

\def\HHtntn{\mathrm{\Hp\Hm \ra \tntn}}
\def\HHcstn{\mathrm{\Hp\Hm \ra \cstn}}
\def\HHcscs{\mathrm{\Hp\Hm \ra \cscs}}

\def\Htn{\mathrm{H^\pm\ra\tau\nu}}
\def\BRTN{\mathrm{Br(H^\pm\ra\tau\nu)}}
\def\MHPM{m_\mathrm{\Hpm}}

\begin{document}

\begin{titlepage}

\mathversion{bold}
\title{Search for Charged Higgs Bosons at LEP}
\mathversion{normal}

\author{The L3 Collaboration}

\begin{abstract}
  
A search for pair-produced charged Higgs bosons is performed with the
L3 detector at LEP using data collected at centre-of-mass energies
between 189 and 209\,\GeV{}, corresponding to an integrated luminosity
of 629.4\,\pb.  Decays into a charm and a strange quark or into a tau
lepton and its neutrino are considered.  No significant excess is
observed and lower limits on the mass of the charged Higgs boson are
derived at the 95\% confidence level. They vary from 76.5 to
82.7\,\GeV, as a function of the $\Htn$ branching ratio.

\end{abstract}

\submitted

\end{titlepage}


\section*{Introduction}

In the Standard Model of the electroweak
interactions~\cite{standard_model} the masses of
bosons and fermions are explained by the Higgs
mechanism~\cite{higgs_mech}. This implies the existence of one
doublet of complex scalar fields which, in turn, leads to a single
neutral scalar Higgs boson.  To date, this Higgs boson has not been
directly observed~\cite{higgs,lnq}.  Some extensions to the Standard
Model contain more than one Higgs doublet~\cite{higgs_hunter}, and
predict Higgs bosons which can be lighter than the Standard Model one
and accessible at LEP.  In particular, models with two complex Higgs
doublets predict two charged Higgs bosons, $\Hpm$, which can be
pair-produced in $\epem$ collisions.

The charged Higgs boson is expected to decay through $\rm\Hp\ra
c\bar{s}$ or $\rm\Hp\ra\tau^+\nu_{\tau}$\footnote{The inclusion of the
charge conjugate reactions is implied throughout this Letter.}, with a
branching ratio which is a free parameter of the models. The process
$\mathrm{\ee\ra\Hp\Hm}$ gives then rise to three different signatures:
$\cscs$, $\cstn$ and $\tntn$. These experimental
signatures have to be disentangled from the large 
background of the $\mathrm{\ee\ra W^+W^-}$ process, characterised by
similar final states.

Data collected at centre-of-mass energies $\sqrt{s}=189-209\,\GeV{}$
are analysed here, superseding  
previous results~\cite{chhiggs_189chhiggs_202}. Data from
$\sqrt{s}=130-183\,\GeV{}$~\cite{chhiggs_130_183} are included to
obtain the final results.  Results from other LEP experiments are given in
Reference~\citen{other_lep}.

The analyses do not depend of flavour tagging variables
and are separately optimised for each of the three possible
signatures.


\section*{Data and Monte Carlo Samples}

The search for pair-produced charged Higgs bosons is performed using
629.4\,\pb{} of data collected in the years from 1998 to 2000 with the
L3 detector~\cite{l3_det} at LEP, at several average centre-of-mass energies,
detailed in Table~\ref{table:lumi}.

The charged Higgs cross section is calculated using the HZHA Monte
Carlo program~\cite{hzha}. As an example, at $\sqrt{s}=206\,\GeV$ it
varies from 0.28\,pb for a Higgs mass, $\MHPM$, of 70\,\GeV{} to
0.17\,pb for $\MHPM=80\,\GeV{}$.  To optimise selections and calculate
efficiencies, samples of $\mathrm{\ee\ra\Hp\Hm}$ events are generated
with the PYTHIA Monte Carlo program~\cite{jetset73} for $\MHPM$
between 50 and 100\,\GeV{}, in steps of 5\,\GeV{}, and between 75 and
80\,\GeV{}, in steps of 1\,\GeV{}.  About 1000 events for each final
state are generated at each Higgs mass.  For background studies, the
following Monte Carlo generators are used: KK2f~\cite{KK2f} for
$\ee\ra\qqbar(\gamma)$, $\ee\ra\mu^+\mu^-(\gamma)$ and
$\ee\ra\tau^+\tau^-(\gamma)$, BHWIDE~\cite{BHWIDE} for $\ee\ra\ee$,
PYTHIA for $\ee\ra\Zo\Zo$ and $\ee\ra\Zo\ee$, YFSWW~\cite{YFSWW} for
$\ee\ra\Wp\Wm$ and PHOJET~\cite{PHOJET} and DIAG36~\cite{DIAG36} for
hadron and lepton production in two-photon interactions, respectively.
The L3 detector response is simulated using the GEANT
program~\cite{my_geant} which takes into account the effects of energy
loss, multiple scattering and showering in the detector.
Time-dependent detector inefficiencies, as monitored during the data
taking period, are included in the simulations.


\section*{Data Analysis}

The analyses for all three final states are updated since our
previous publications at lower centre-of-mass
energies~\cite{chhiggs_189chhiggs_202,chhiggs_130_183}: the searches
in the $\HHcscs$ and $\cstn$ channels are based on a mass dependent
likelihood interpretation of data samples selected~\cite{sigmaW}
for the studies of W pair-production, while a discriminant variable is
introduced for the search in the $\tntn$ channel. These analyses are
described below.  

\subsection*{{\boldmath{Search in the $ \HHcscs$ channel}}}

The search in the $\HHcscs$ channel proceeds from a selection of high
multiplicity events with balanced transverse and longitudinal momenta
and with a visible energy which is a large fraction of
$\sqrt{s}$. These criteria reject events from low-multiplicity
processes like lepton pair-production, events from two-photon
interactions and pair-production of W bosons where at least one boson decays into leptons.
The events are forced into four jets by means of the
DURHAM algorithm~\cite{durham} and a neural network~\cite{sigmaW}
discriminates between events which are compatible with a four-jet
topology and those  from the large cross section
$\ee\ra\qqbar(\gamma)$ process in which four-jet events originate from
hard gluon radiation. The neural network inputs are the event
spherocity, the energies and widths of the most and least energetic
jets, the difference between the energies of the second and third most
energetic jets, the minimum multiplicity of calorimetric clusters and charged tracks for
any jet, the value of the $y_{34}$ parameter of the DURHAM algorithm
and the compatibility with energy-momentum conservation in
$\epem$ collisions. After a cut on the output of the neural network,
two constrained fits are performed. The first four-constraint fit
enforces energy and momentum conservation, modifying the jet energies
and directions. The second five-constraint (5C) fit imposes the
additional constraint of the production of two equal mass
particles. Among the three possible jet pairings, the one is retained
which is most compatible with this equal mass hypothesis.  Events with
a low probability for the fit hypotheses are removed from the sample
and a total of 5156 events are observed in data while 5112 are
expected from Standard Model processes. The corresponding signal
efficiencies are between 70\% and 80\%, for $\MHPM = 60-95\,\GeV$.

Likelihood variables~\cite{mssm} are built to discriminate four-jet
events compatible with charged Higgs production from the dominating
background from W pair-production. A different likelihood is prepared
for each simulated Monte Carlo sample corresponding to a different
Higgs boson mass. Seven variables are included in the likelihoods:

\begin{itemize}
\item the minimum opening angle between  paired jets;
\item the difference between the largest and smallest jet energies;
\item the difference between the di-jet masses;
\item the output of the neural network for the selection of four-jet
  events;
\item the absolute value of the cosine of the polar angle of the thrust vector;
\item the cosine of the polar angle at which the positive
  charged\footnote{Charge assignment is based on jet-charge
  techniques~\cite{TGC}.} boson is produced; 
\item the value of the quantity $2\ln|M|$, where $M$ is the matrix
  element for the $\epem\ra{\rm W^+W^-}\ra four fermions$ process from
  the EXCALIBUR~\cite{EXCALIBUR} Monte Carlo program, calculated using
  the four-momenta of the reconstructed jets.
\end{itemize}

Figures 1a, 1b and 1c show the distributions of the last three
variables while Figure~1d presents the distribution of the
likelihood variable for $\MHPM = 70\,\GeV$. A cut at 0.7 on this
variable, 
which maximizes the signal sensitivity, is applied as a final selection
criterion, for all mass hypotheses.  The numbers of
observed and expected events are given in
Table~\ref{events} and the selection efficiencies in
Table~\ref{eff}. The main contributions to the background come from
hadronic W-pair decays (70\%) and from the
$\ee\ra\qqbar(\gamma)$ process (26\%).  Figure~2 shows the 5C mass of the
pair-produced bosons before and after the cut on the final likelihoods.
Peaks from pair-production of W as well as Z bosons are visible.

\subsection*{{\boldmath{Search in the $\HHcstn$ channel}}}

The search in the $\HHcstn$ channel selects events with high multiplicity, two
hadronic jets and a tau candidate. Tau candidates can be identified
either as electrons or muons with momentum incompatible with that
expected for leptons originating from direct semileptonic decay of W
pairs, or with narrow, low multiplicity jets with at least one charged
track, singled out from the hadronic background with a neural
network~\cite{sigmaW}.  The tau energy is reconstructed by imposing
four-momentum conservation and enforcing the hypothesis of the
production of two equal mass particles.  The events must have a
transverse missing momentum of at least 20\,\GeV{} and the absolute
value of the cosine of the polar angle of the missing momentum is
required to be less than 0.9. Finally, the di-jet invariant mass is
required to be less than 100\,\GeV{} and the mass recoiling against the
di-jet system less than 130\,\GeV, thus selecting 1026 events in data
while 979 are expected from Standard Model processes, mainly from W
pair-production where one of the W bosons decays into leptons and the
other into hadrons. The signal efficiency is about 50\%.

To discriminate the signal from the background,  mass dependent
likelihoods~\cite{mssm} are built which contain eight variables:

\begin{itemize}

\item the di-jet acoplanarity;
\item the angle of the tau flight direction with respect to that of
      its parent boson in the rest frame of the latter;
\item the di-jet mass;
\item the quantity $2\ln|M|$ calculated using the four-momenta of the
  reconstructed jets and tau as well as the missing momentum
  and energy;
\item the transverse momentum of the event, normalised to $\sqrt{s}$;
\item the polar angle of the hadronic system, multiplied
  by the charge of the reconstructed tau;
\item the sum $\Sigma_\theta$ of the angles between the tau candidate and the nearest
  jet and between the missing momentum and the nearest jet;
\item the energy of the tau candidate, calculated in the rest frame of its parent
  boson and scaled by $\sqrt{s}$.
\end{itemize}

The distributions of the last three variables are shown in Figures~3a,
3b and 3c. Figure~3d presents an example of the distributions of the
likelihood variable for $\MHPM=70\,\GeV$ for data, background and
signal Monte Carlo. A cut at 0.6 is applied for all likelihoods. This
cut corresponds to the largest sensitivity to a charged Higgs signal.
Table~\ref{events} gives the numbers of observed and expected events,
while the selection efficiencies are given in Table~{\ref{eff}}. Over
95\% of the background is due to W pair-production. Figure~4 shows the
reconstructed mass of the pair-produced bosons before and after the
cut on the final likelihoods.

\subsection*{{\boldmath{Search in the $\HHtntn$ channel}}}

The signature for the leptonic decay channel is a pair of tau
leptons. These are identified either via their decay into electrons or
muons, or as narrow jets.

The selection criteria are similar to those used at lower
$\sqrt{s}$~\cite{chhiggs_189chhiggs_202,chhiggs_130_183}. Low
multiplicity events with large missing energy and momentum are
retained. To reduce lepton-pair background, an upper cut is placed on
the value of the event collinearity angle, $\xi$, defined as the maximum angle
between any pair of tracks. The distribution of this variable is shown
in Figure~5a.  The contribution from cosmic muons is reduced by making use of
information from the time-of-flight system. Figure~\ref{fig:lepton}b
presents the distribution of the scaled visible energy, $E_{vis}/\sqrt{s}$, for
events on which all other selection criteria are applied.

The analysis is modified with respect to those previously
published~\cite{chhiggs_189chhiggs_202,chhiggs_130_183} in that the
normalised transverse missing momentum of the event, ${P_t/E_{vis}}$, whose
distribution is shown in Figure~\ref{fig:lepton}c, is
used as a linear discriminant variable on which no cut is applied.

The efficiency of the $\HHtntn$ selection for several Higgs masses is
listed in Table~\ref{eff}.  The numbers of observed and expected events
are presented in Table~\ref{events}.  The background is mainly formed
by W-pair production (60\%), two-photon interactions (26\%) and lepton pair-production (9\%).


\section*{Results}

The number of selected events in each decay channel is consistent with
the number of events expected from Standard Model processes.  A
technique based on a log-likelihood ratio~\cite{lnq} is used to
calculate a confidence level (CL) that the observed events are
consistent with background expectations.  For the $\cscs$ and $\cstn$
channels, the reconstructed mass distributions, shown in
Figures~\ref{fig:cscs_mass}b and~\ref{fig:cstn_mass}b, are used in the
calculation, whereas for the $\tntn$ channel, the distribution of the
normalised transverse missing momentum, shown in
Figure~\ref{fig:lepton}c, is used.

The systematic uncertainties on the background level and the signal
efficiencies are included in the confidence level calculation. These
are due to finite Monte Carlo statistics and to the uncertainty on the
background normalisation. The former uncertainty is 5\% for
the background and 2\% for the signal Monte Carlo samples. The
uncertainty on the background normalisation is 3\% for the $\HHcscs$
channel and 2\% for the $\cstn$ and $\tntn$ channels.  The systematic
uncertainty on the signal efficiency due to the selection procedure is
estimated by varying the selection criteria and is found to be less
than 1\%.  These systematic uncertainties decrease the $\MHPM$
sensitivity of the combined analysis by about 200\,\MeV.

Figure~\ref{fig:clb} compares the resulting background confidence
level, $1-{CL_b}$, for the data to the expectation in the absence of a
signal, for three values of the $\Htn$ branching ratio: $\BRTN$ = 0,
0.5 and~1. The 68.3\% and 95.4\% probability bands expected in the
absence of a signal are also displayed and denoted as $1\sigma$ and
$2\sigma$, respectively. A slight excess of data appears around $\MHPM
= 69$\,\GeV{} for $\BRTN = 0$, as previously
observed~\cite{chhiggs_189chhiggs_202}. It is compatible with a
$2.5\sigma$ upward fluctuation in the background. The excess is
also compatible with a $2.9\sigma$
downward fluctuation of the signal\footnote{As an example, for $\BRTN = 0.1$, these figures are
$1.8\sigma$ and $2.7\sigma$, respectively.}.
As observed in Figures 6b and 6c, no excess is present in the $\cstn$
and $\tntn$ channels around $\MHPM = 69$\,\GeV{}. Therefore, the $\cscs$
excess is interpreted as a statistical fluctuation in the background
and lower limits at the 95\% CL on $\MHPM$ are derived~\cite{lnq} as a
function of $\BRTN$. Data at $\sqrt{s}=130-183\,\GeV$~\cite{chhiggs_130_183} are included to
obtain the limits.  Figure~\ref{exclusion} shows the excluded $\MHPM$
regions for each of the final states and their combination, as a
function of $\BRTN$.  Table~\ref{cl} gives the observed and the
median expected lower limits for several values of the branching
ratio.

In conclusion, refined analyses and larger centre-of-mass energies
improve the sensitivity of the search for charged Higgs bosons produced in $\epem$
collisions as compared to previous
results~\cite{chhiggs_189chhiggs_202,chhiggs_130_183}. No
significant excess is observed in data and a lower limit at 95\% CL on
the charged Higgs boson mass is obtained as $$\MHPM > 76.5\,\GeV{},$$
independent of its branching ratio. 


\bibliographystyle{l3stylem}

\newpage

%
\newpage
\section*{Author List}
\input namelist274.tex

\begin{table}
\begin{center}
\begin{tabular}{|l|cccccccc|}
\hline
$\sqrt{s}$ (\GeV)      & 188.6 & 191.6 & 195.5 & 199.5 & 201.7 & 204.9 & 206.4 & 208.0 \\
Luminosity (pb$^{-1}$) & 176.8 & \phantom{0}29.8 &  \phantom{0}84.2 &  \phantom{0}83.3 &  \phantom{0}37.2 &  \phantom{0}79.0 & 130.8 &   \phantom{00}8.3 \\
\hline
\end{tabular}
\caption{Average centre-of-mass energies and corresponding integrated luminosities.}
\label{table:lumi}
\end{center}
\end{table}

\begin{table}
\begin{center}
\begin{tabular}{|l|ccc|} \cline{2-4}
\multicolumn{1}{c|}{} & \multicolumn{3}{|c|}{Channel} \\
\multicolumn{1}{c|}{} &  $\cscs$ & $\cstn$  & $\tntn$  \\ \hline
 Data                 &  2296    & 442    & 141\\
 Background           &  2228    & 464    & 141   \\ \hline
 Signal               &   \phantom{0}100    &  \phantom{0}76    &  \phantom{0}50   \\ \hline
\end{tabular}
\caption{\label{events}  Number of observed data  events  and  background
    expectations in the three analysis channels.  The  uncertainty  on
    the background  expectations  is 
    estimated to be 5\%. The numbers of expected signal events for
     $\MHPM=70\,\GeV$ and $\BRTN=0$, 0.5 and 1 are also given for the
     $\cscs$, $\cstn$ and $\tntn$ channels, respectively.}
\end{center}
\end{table}

\begin{table}
\begin{center}
\begin{tabular}{|l|cccccc|r|r|} \hline
\multicolumn{1}{|c|}{\raisebox{-8pt}[0pt][0pt]{Channel}}
         & \multicolumn{6}{c|}{Selection efficiency (\%)} \\
         &$\MHPM=$& 60 \GeV & 70 \GeV & 80 \GeV & 90 \GeV & 95 \GeV \\ \hline
$\cscs$  && 62      & 62      & 50      & 58      & 64      \\
$\cstn$  && 38      & 51      & 43      & 43      & 39      \\
$\tntn$  && 26      & 30      & 33      & 34      & 36      \\ \hline
\end{tabular}
\caption{\label{eff}   Selection  efficiencies  for
    various charged Higgs masses.  The  efficiencies  are largely  independent of
    the  centre-of-mass  energy.  The  uncertainty on each efficiency is
    estimated to be 2\%.}

\end{center}
\end{table}

\begin{table}
\begin{center}
\begin{tabular}{|c|cc|} \hline
\raisebox{-8pt}[0pt][0pt]{$\BRTN$}
      & \multicolumn{2}{|c|}{Lower limits (\GeV{}) at 95\% CL} \\
      & \,\,\,\,\,\,\, observed & expected \\ \hline
 0.0\phantom{0}  & \,\,\,\,\,\,\, 76.7 & 77.5  \\
 0.26            & \,\,\,\,\,\,\, 76.5 & 75.6  \\
 0.5\phantom{0}  & \,\,\,\,\,\,\, 76.6 & 76.5  \\
 1.0\phantom{0}  & \,\,\,\,\,\,\, 82.7 & 84.6  \\ \hline
\end{tabular}
\caption{Observed  and    expected  lower  limits  at 95\%  CL for
    different   values  of  the  $\Htn$  branching  ratio.  The  minimum
    observed limit  is at $\BRTN
    = 0.26$.}
\label{cl}
\end{center}
\end{table}


\newpage

\begin{figure}[hp]
\mbox{\epsfig{figure=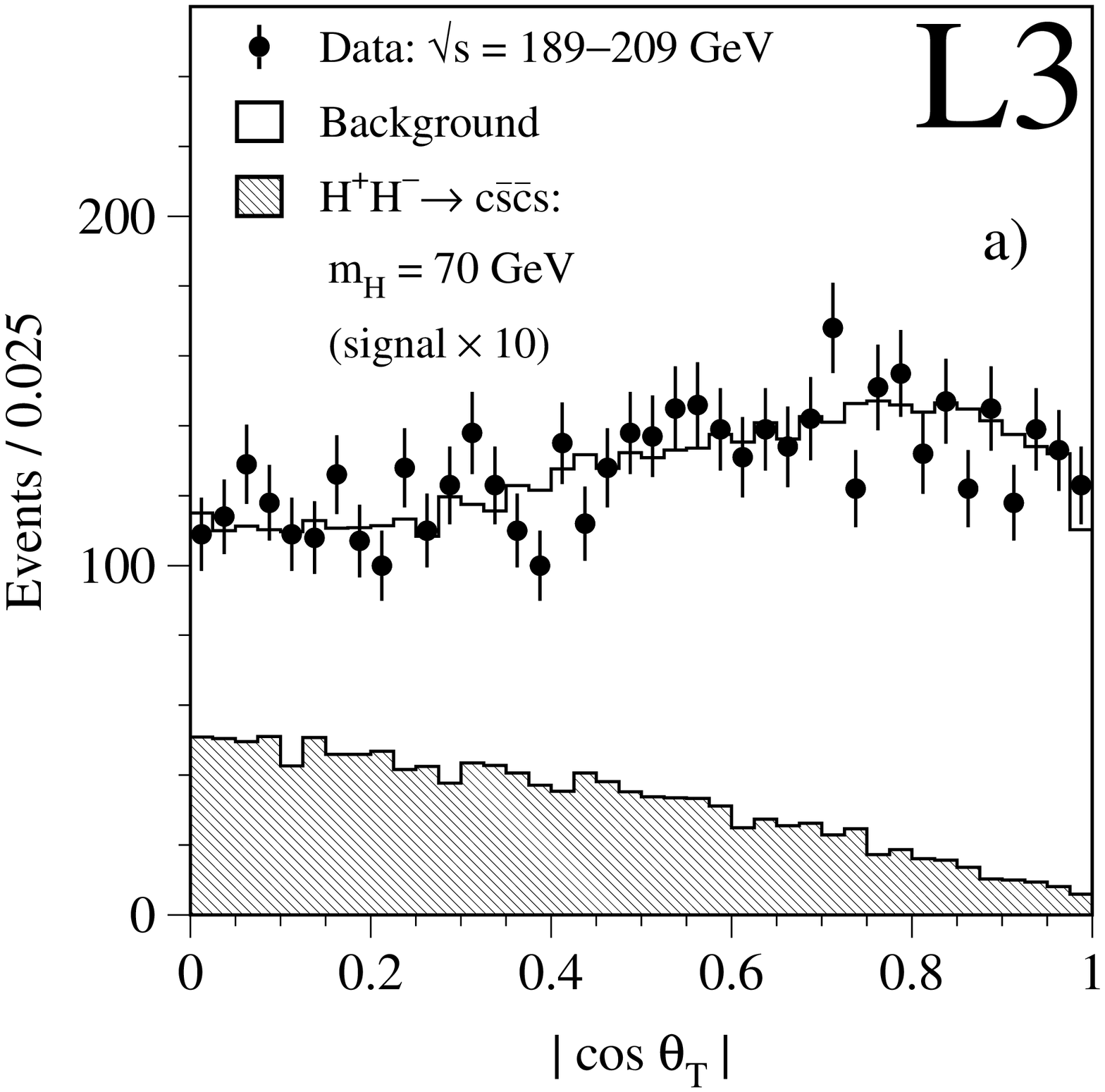,width=0.5\textwidth}
      \epsfig{figure=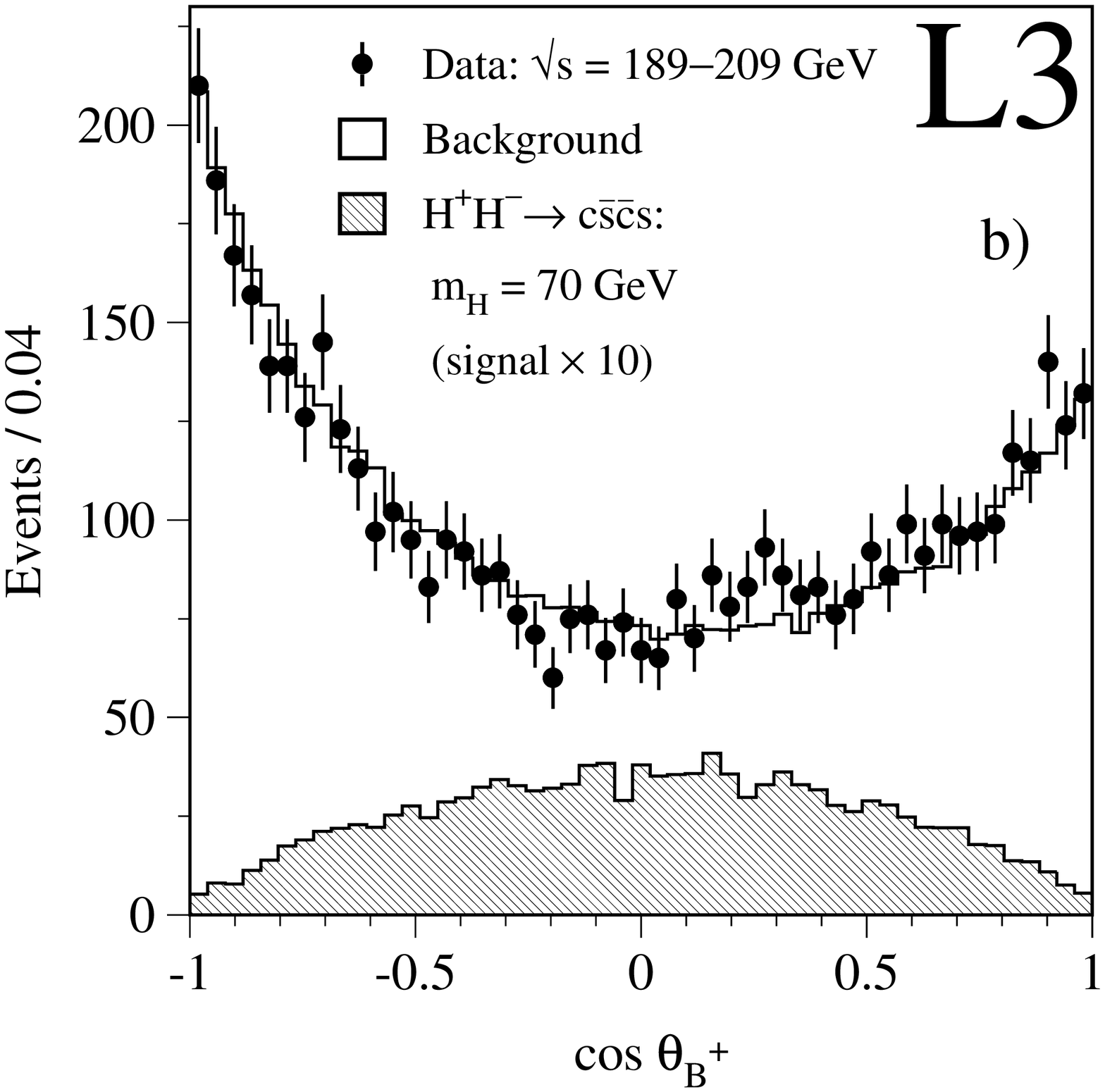, width=0.5\textwidth}}
\mbox{\epsfig{figure=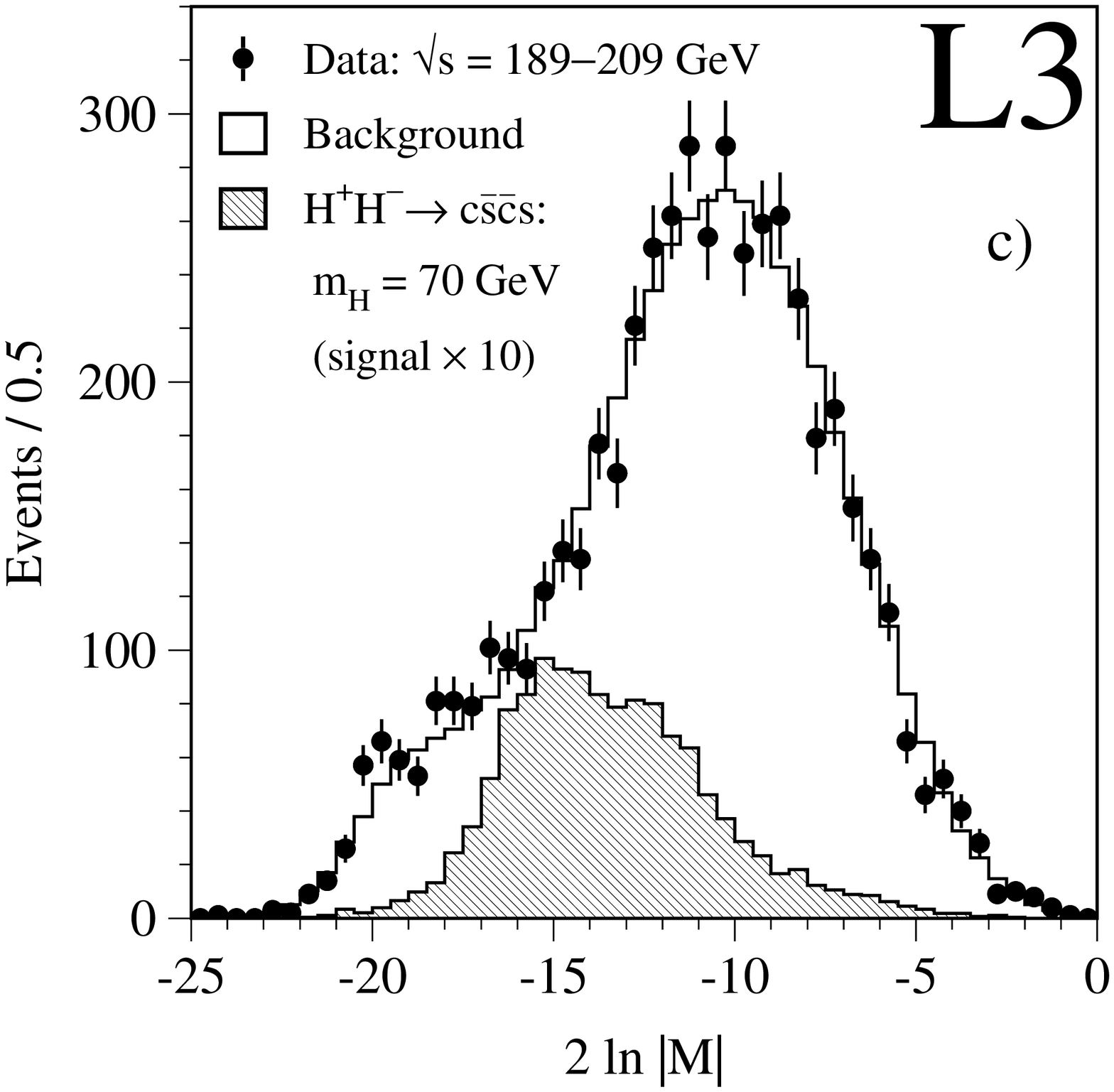,width=0.5\textwidth}
      \epsfig{figure=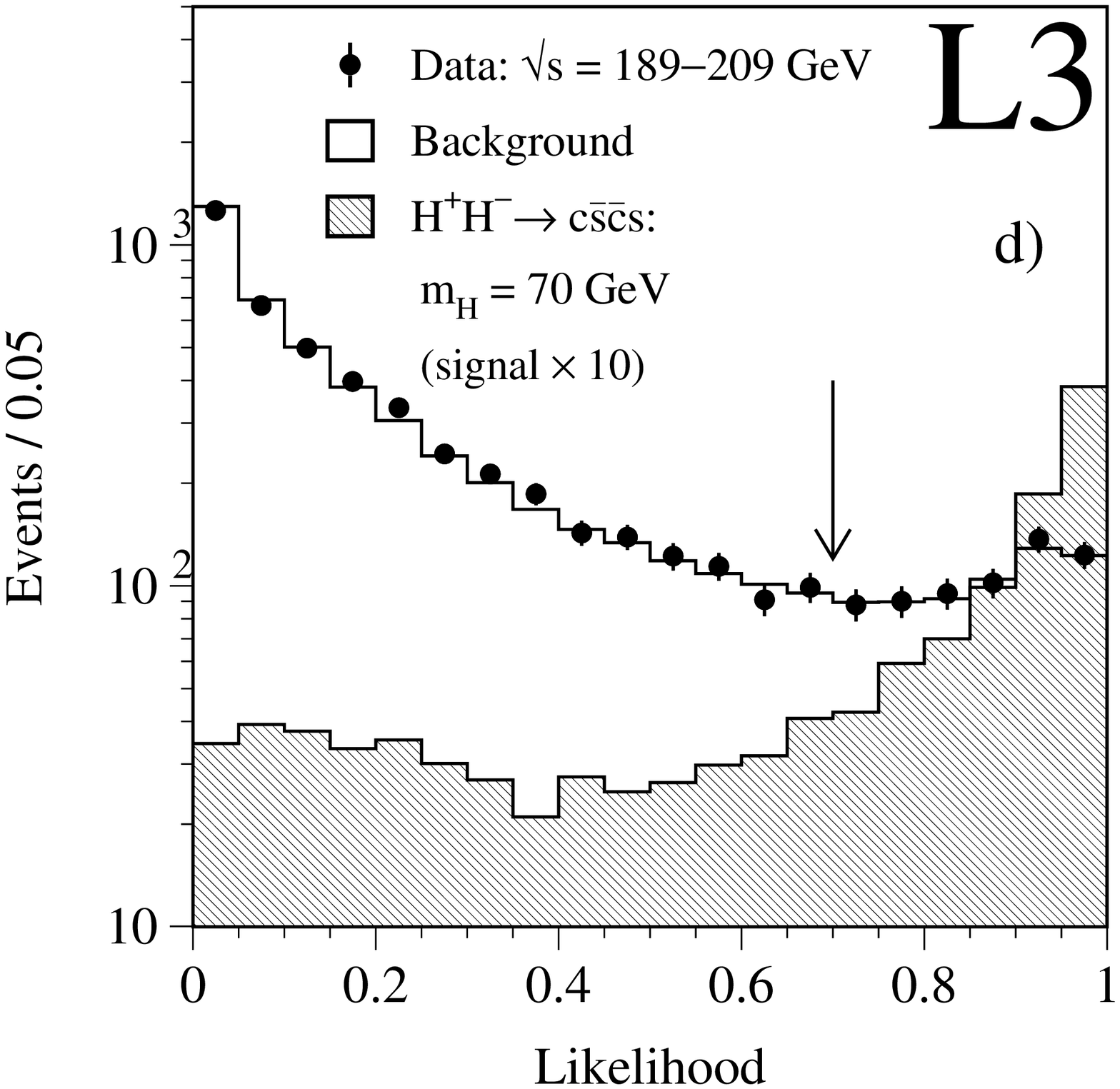,width=0.5\textwidth}}
\caption[]{\label{fig:cscs_var1} Distributions  for the
  $\HHcscs$ channel of: a) the absolute value of the cosine of the polar angle of 
  the thrust axis,  b) the cosine of the polar angle of 
  the positively charged boson, c) the logarithm of the squared matrix element
  for the $\epem\ra\Wp\Wm$ process and d) the selection likelihood for
  $\MHPM=70\,\GeV{}$.  The points represent the data  
  and the open histogram the expected background.  The hatched
  histogram indicates the expected distribution for a signal with
  $\MHPM=70\,\GeV{}$ and $\BRTN = 0$, multiplied by a factor of 10.  The
  arrow in d) shows the position of the cut.}
\end{figure}

\begin{figure}[hp]
\centerline{\epsfig{figure=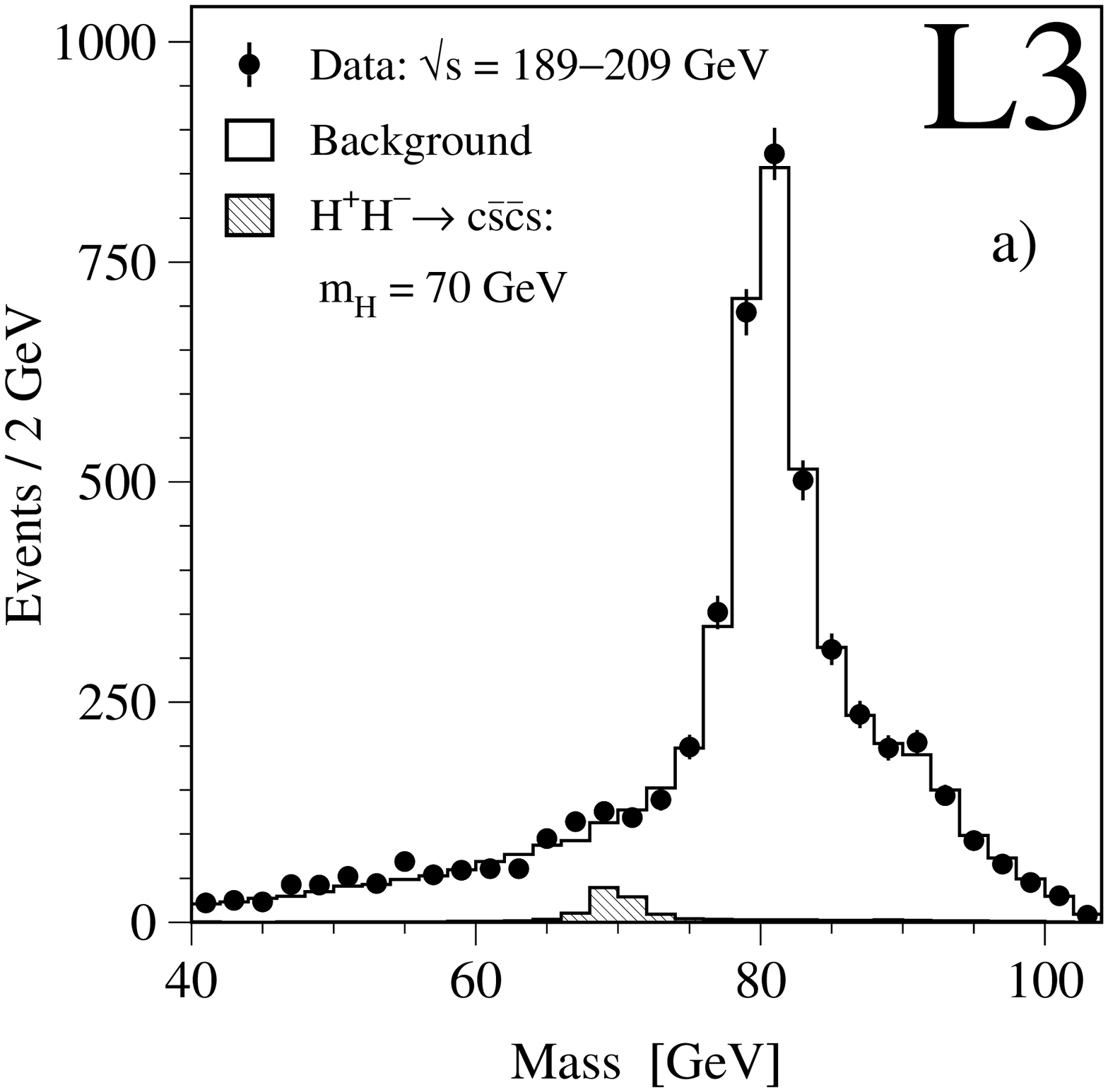,width=0.6\textwidth}}
\centerline{\epsfig{figure=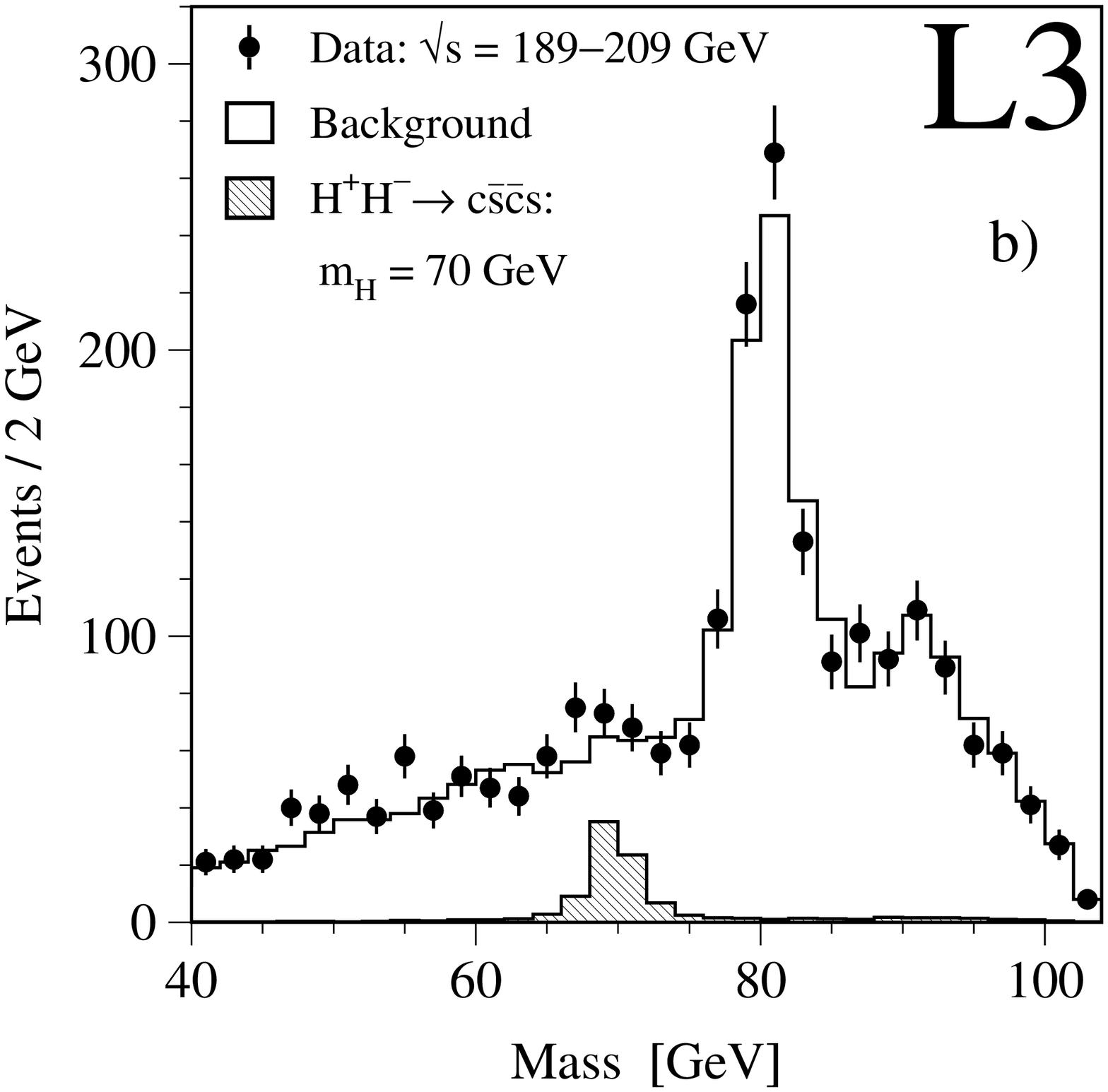,width=0.6\textwidth}}
\caption[]{\label{fig:cscs_mass} Reconstructed mass spectra in the
  $\HHcscs$ channel, for data and expected background, for events a)
  before, and b) after, the cut on the likelihoods.  The points represent the data
  and the open histogram the expected background. The expected
  distribution for $\MHPM=70\,\GeV{}$ and $\BRTN = 0$ is
  shown as the hatched histogram.}
\end{figure}

\begin{figure}[hp]
\mbox{\epsfig{figure=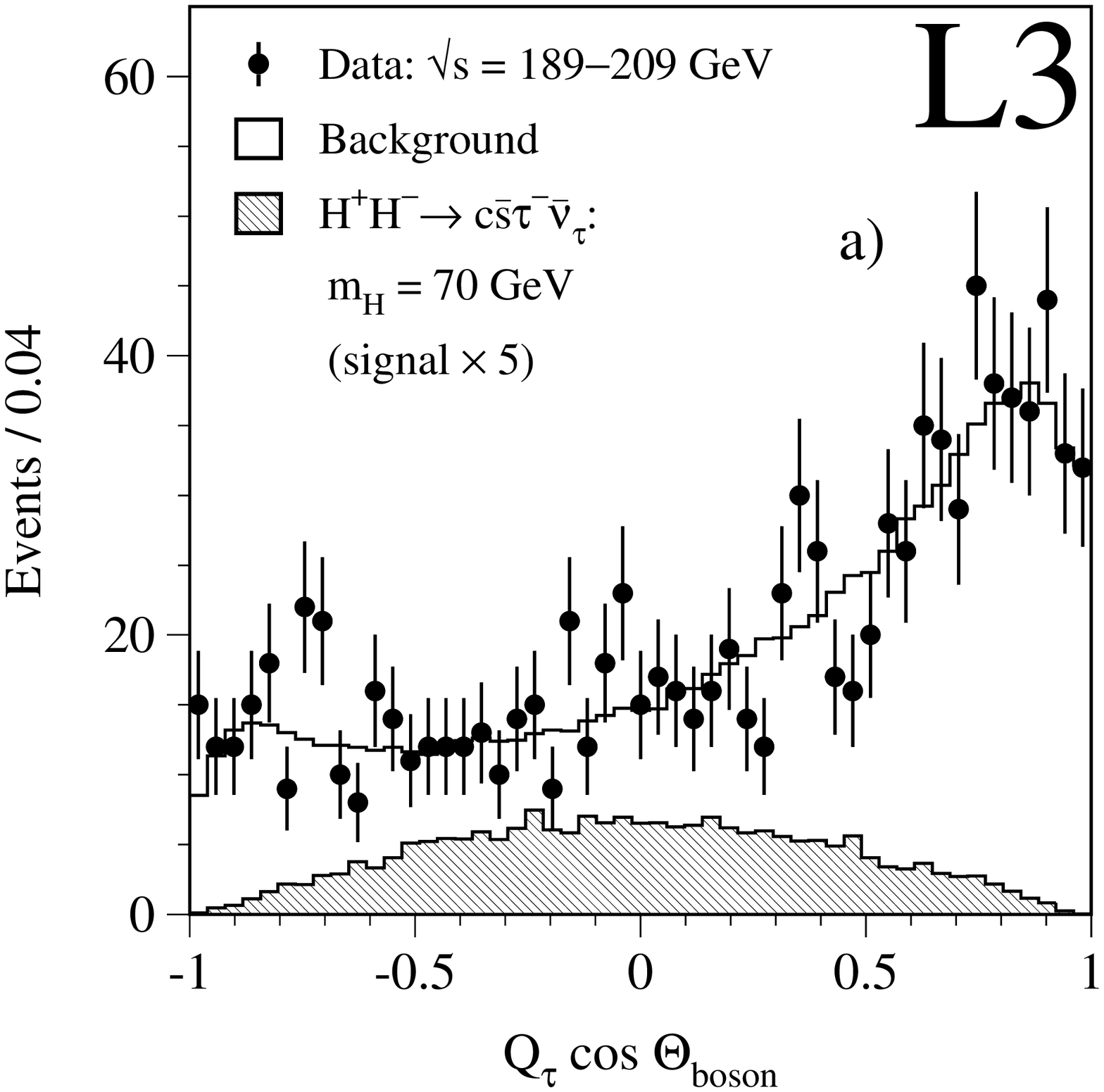,width=0.5\textwidth}%
      \epsfig{figure=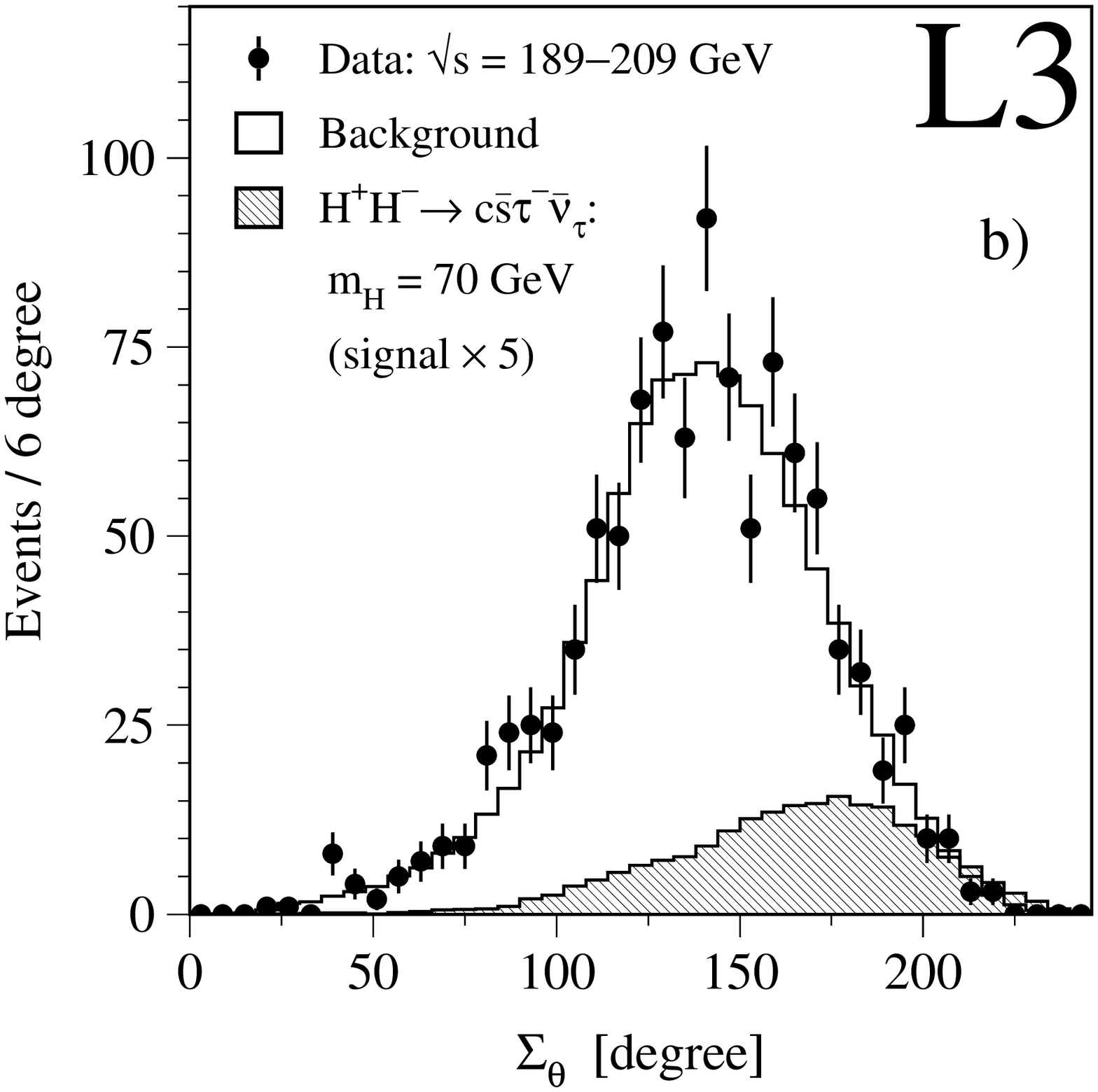,width=0.5\textwidth}}
\mbox{\epsfig{figure=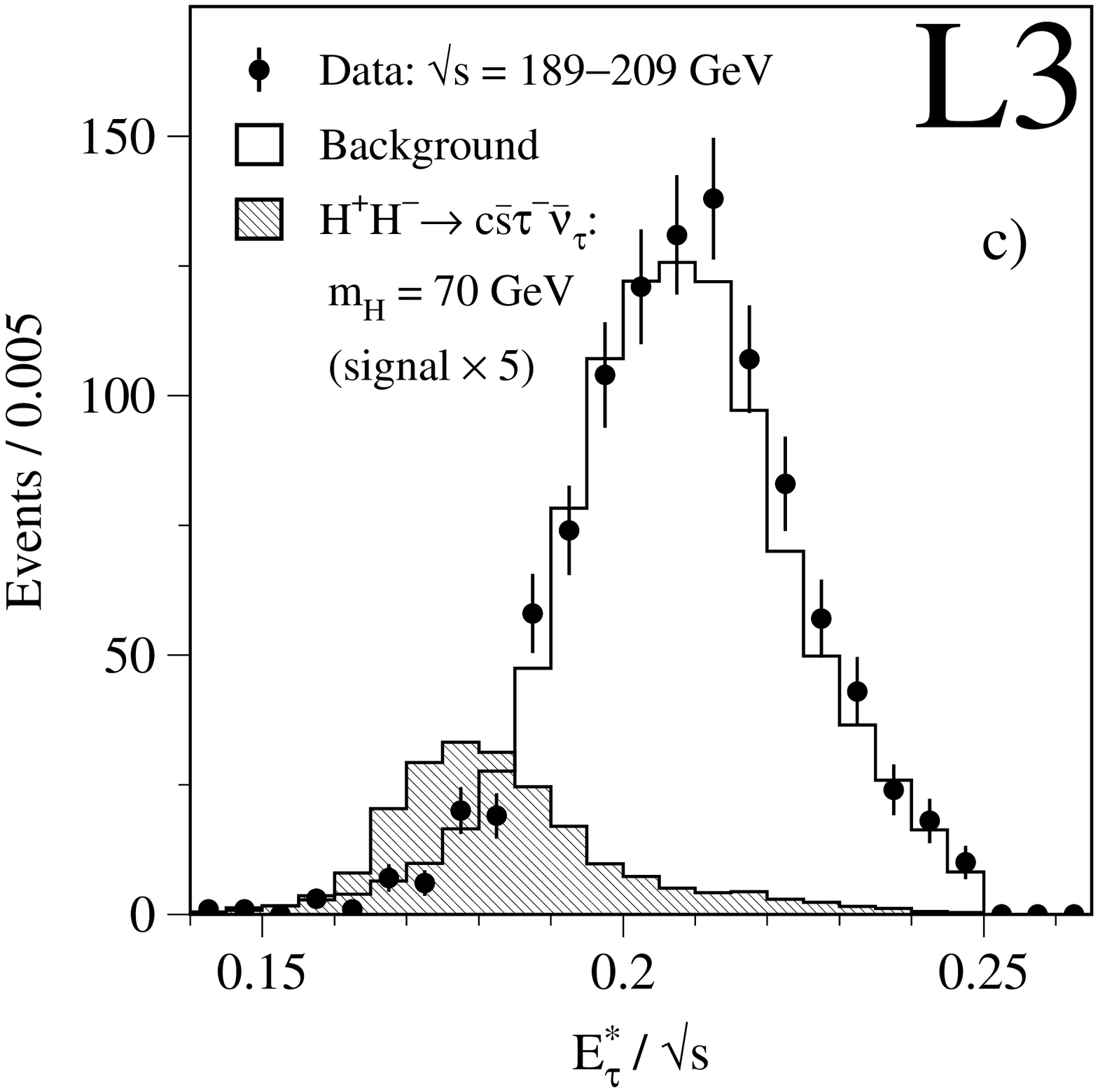,width=0.5\textwidth}%
      \epsfig{figure=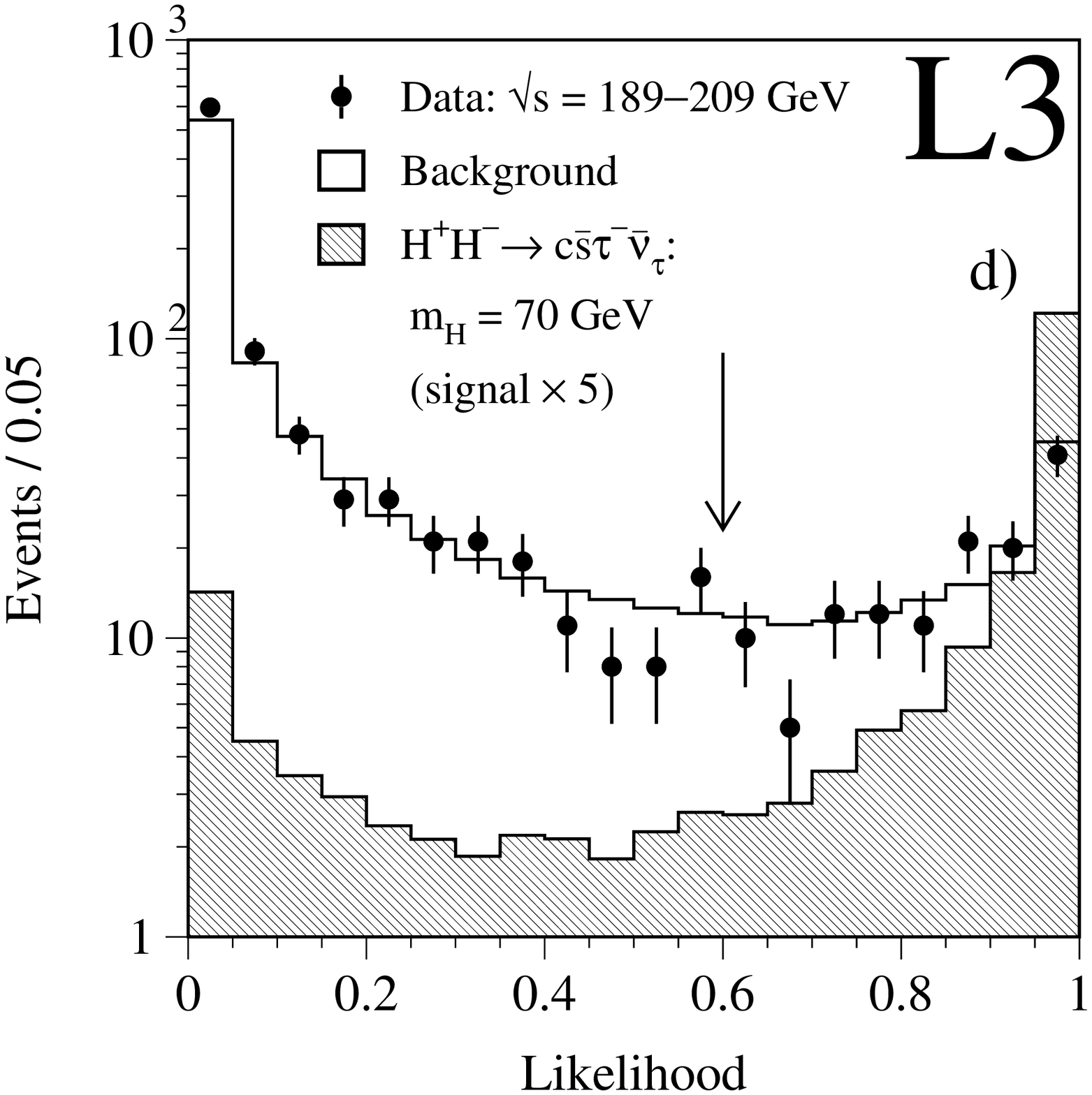,width=0.5\textwidth}}
\caption[]{\label{fig:cstn_var2} Distribution for the
  $\HHcstn$ channel of: a) the cosine of the polar angle of the hadron
  system multiplied by the charge of the tau candidate, b) the sum of
  the angles between the tau candidate and the nearest jet and between
  the missing momentum and the nearest jet, c) the scaled energy of
  the tau candidate in the rest frame of the parent boson and d) the
  selection likelihood for $\MHPM=70\,\GeV{}$.  The points represent
  the data and the open histogram the expected background.  The
  hatched histogram indicates the expected distribution for
  $\MHPM=70\,\GeV{}$ and $\BRTN = 0.5$, multiplied by a factor of
  5. The arrow in d) shows the position of the cut.}
\end{figure}

\begin{figure}[hp]
\centerline{\epsfig{figure=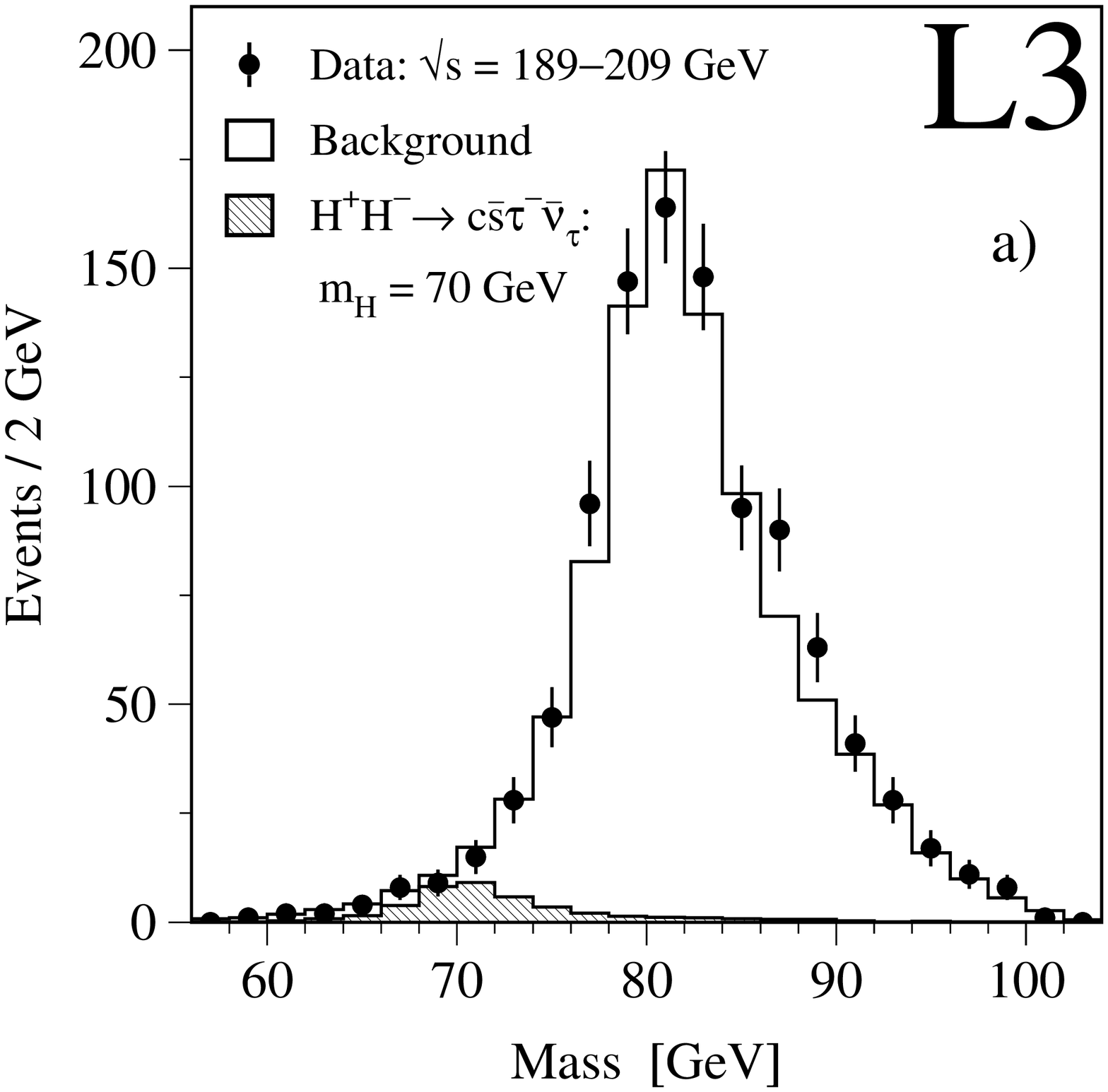,width=0.6\textwidth}}
\centerline{\epsfig{figure=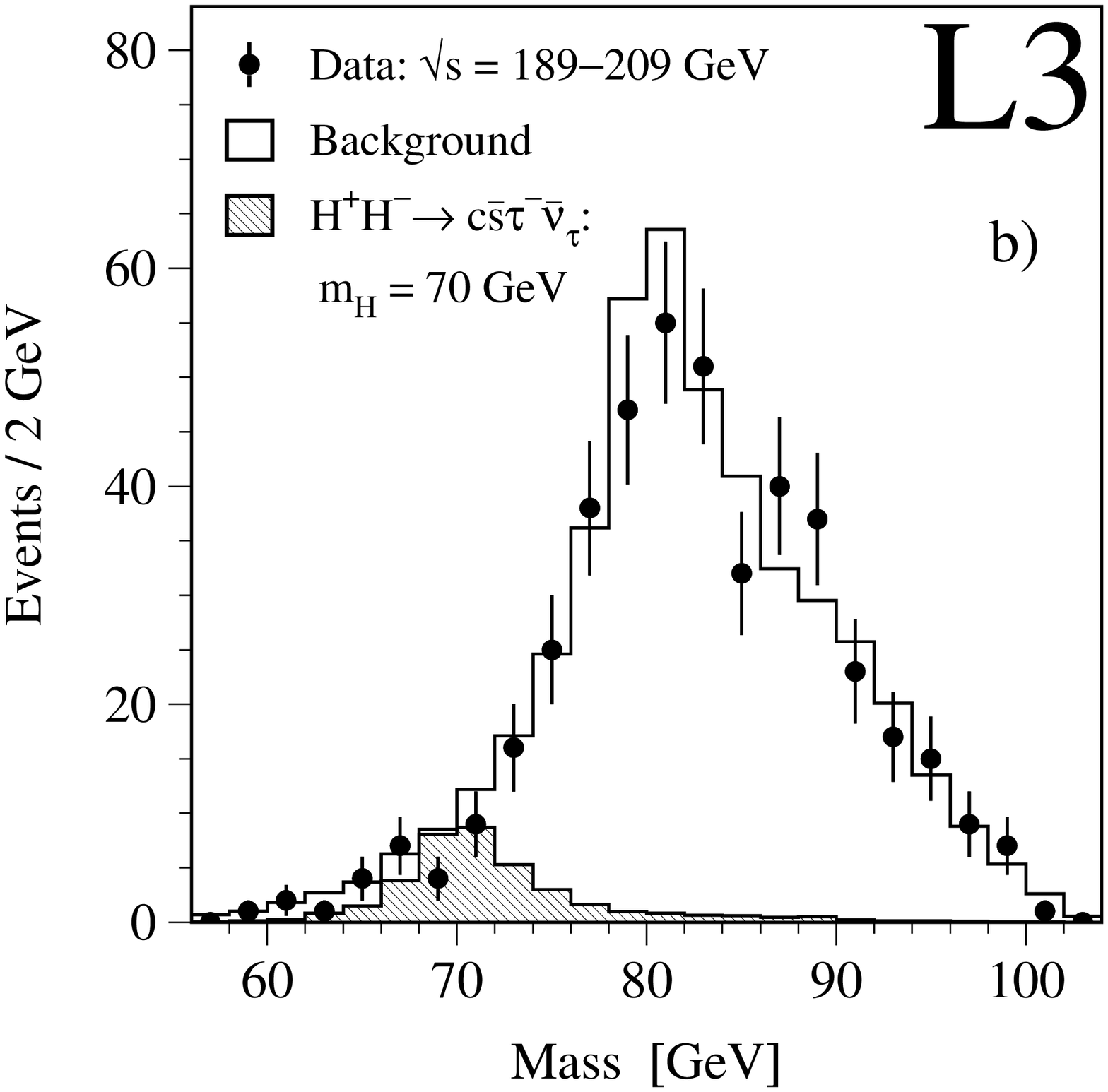,width=0.6\textwidth}}
\caption[]{\label{fig:cstn_mass} Reconstructed mass spectra in the
  $\HHcstn$ channel, for data and expected background, for events a)
  before, and b) after, the cut on the likelihoods.  The points represent the data
  and the open histogram the expected background. The expected
  distribution for  $\MHPM=70\,\GeV$ and $\BRTN = 0.5$ is shown as
  the hatched histogram.}
\end{figure}

\begin{figure}[hp]
\mbox{\epsfig{figure=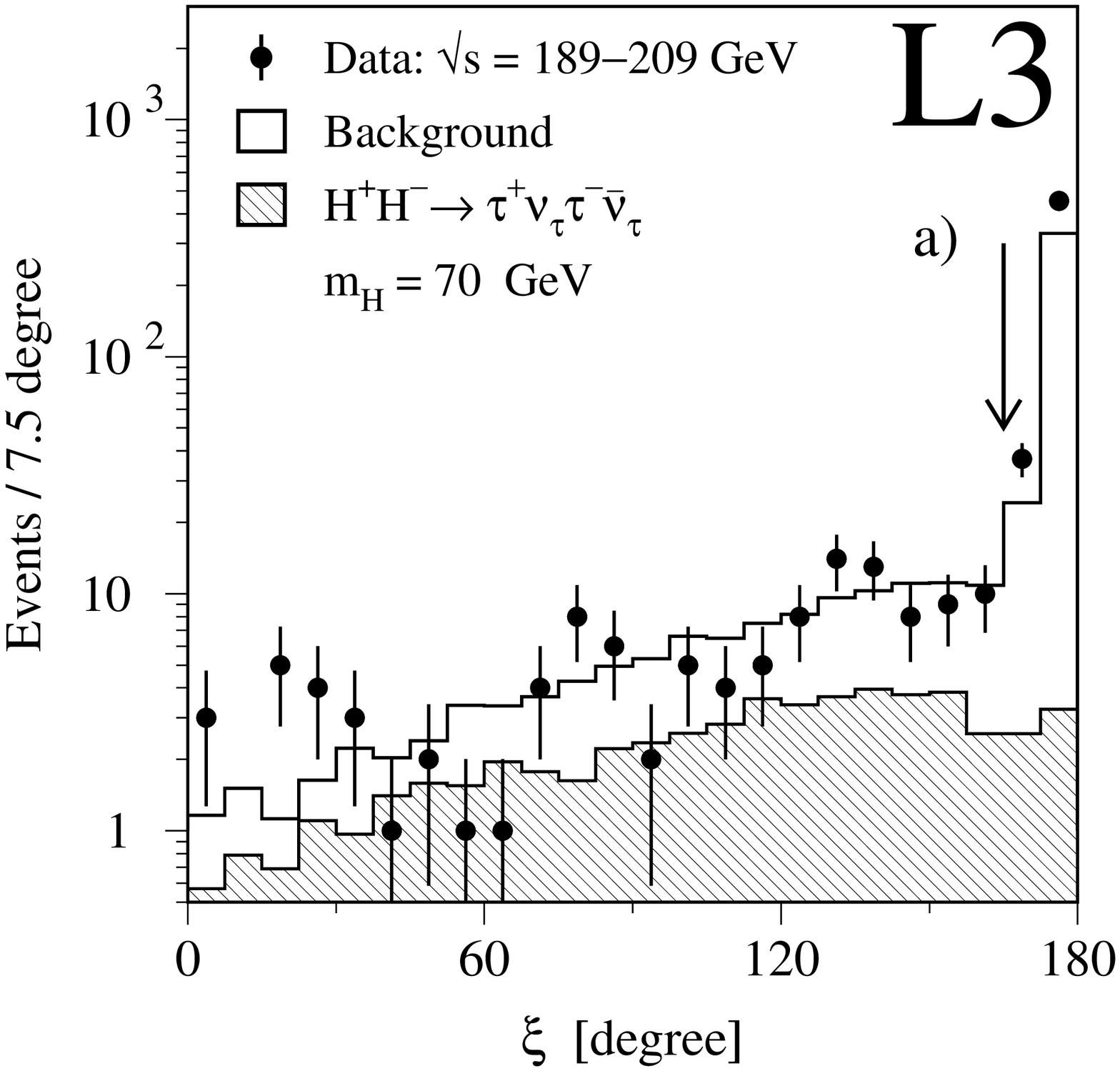,width=0.5\textwidth}%
      \epsfig{figure=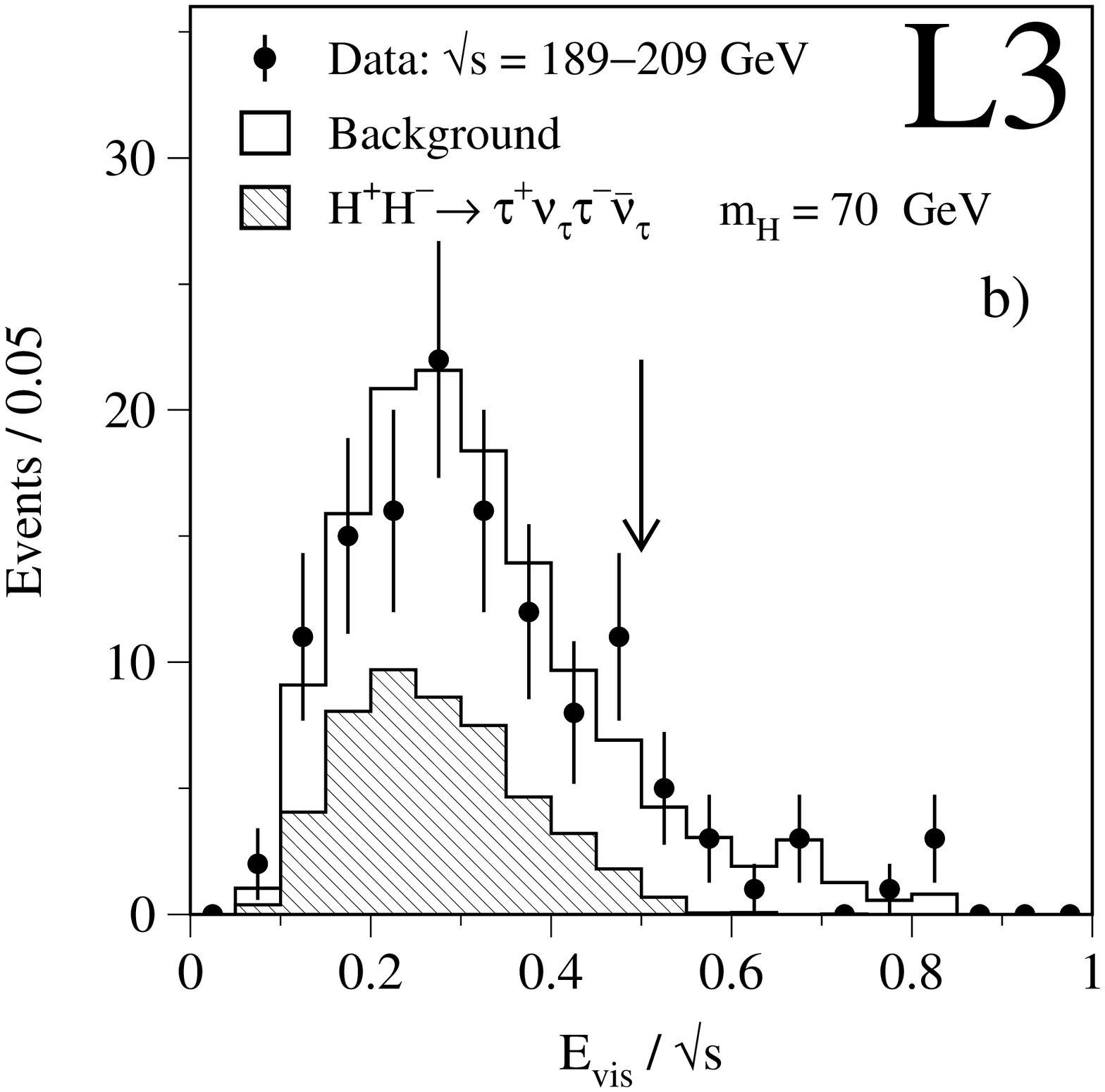,width=0.5\textwidth}}
\centerline{\epsfig{figure=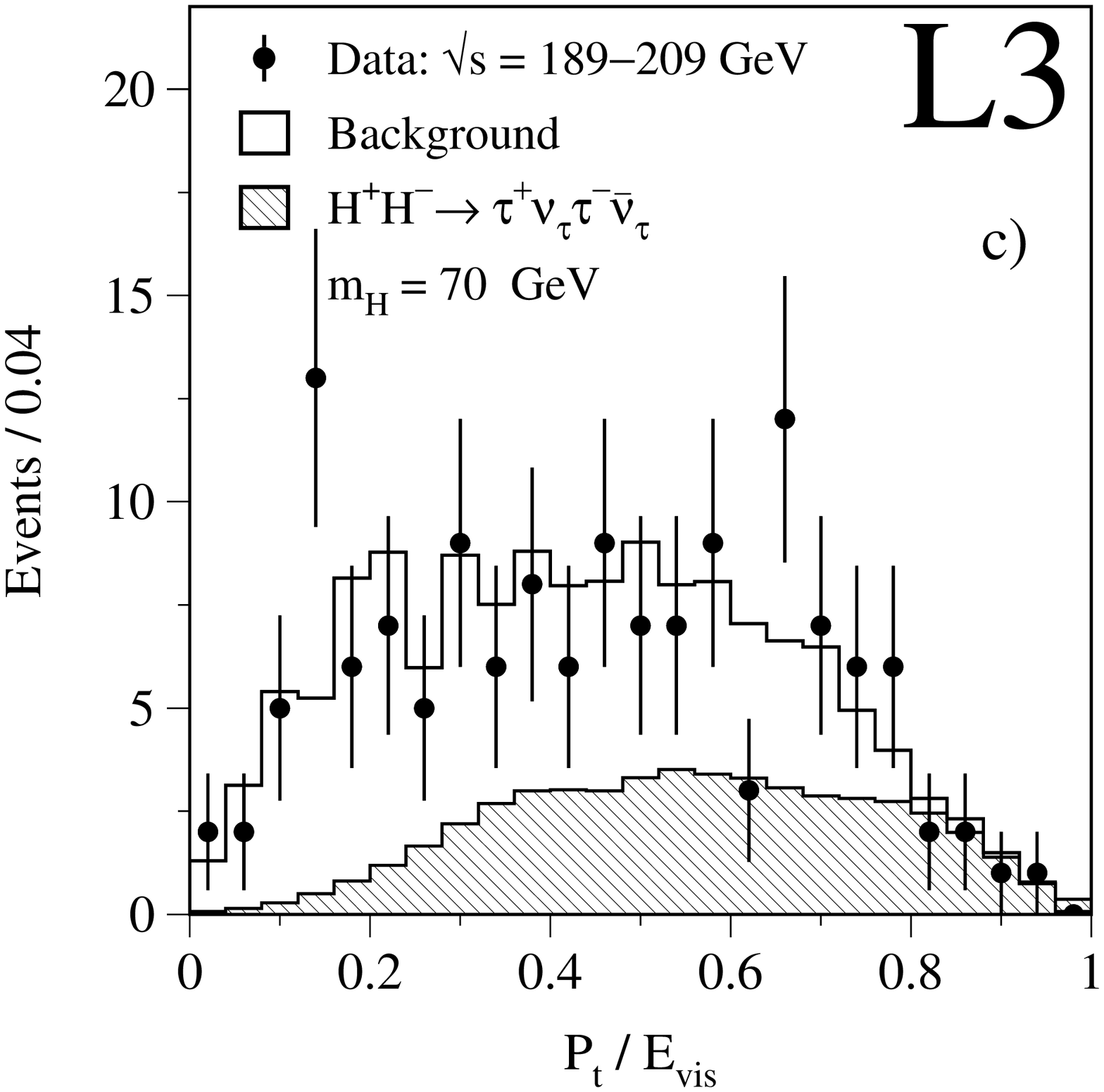,width=0.5\textwidth}}
\caption[]{\label{fig:lepton} Distribution for the $\HHtntn$ channel
  of: a) the event collinearity angle, $\xi$, b) the scaled visible energy and c) the
  normalised transverse missing momentum of the event. In a) and b) all other
  selection criteria are applied and the arrows indicate the cut on
  the displayed variable. The points represent the data and the open
  histogram the expected background.  The hatched histograms indicate
  the expected signal distributions for 
  $\MHPM=70\,\GeV{}$ and $\BRTN = 1$.}
\end{figure}

\begin{figure}[hp]
      \mbox{\epsfig{figure=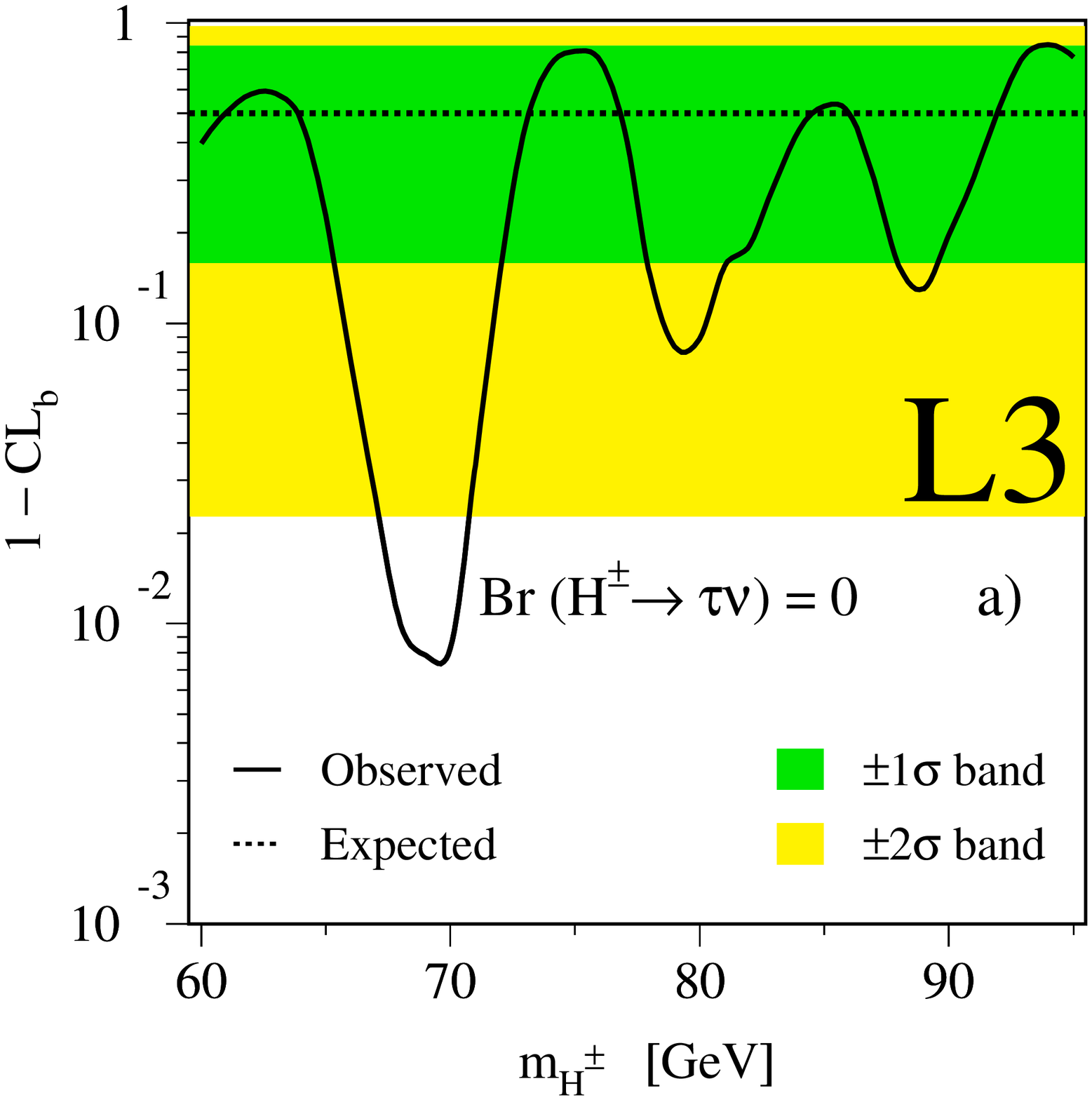,width=0.5\textwidth}%
            \epsfig{figure=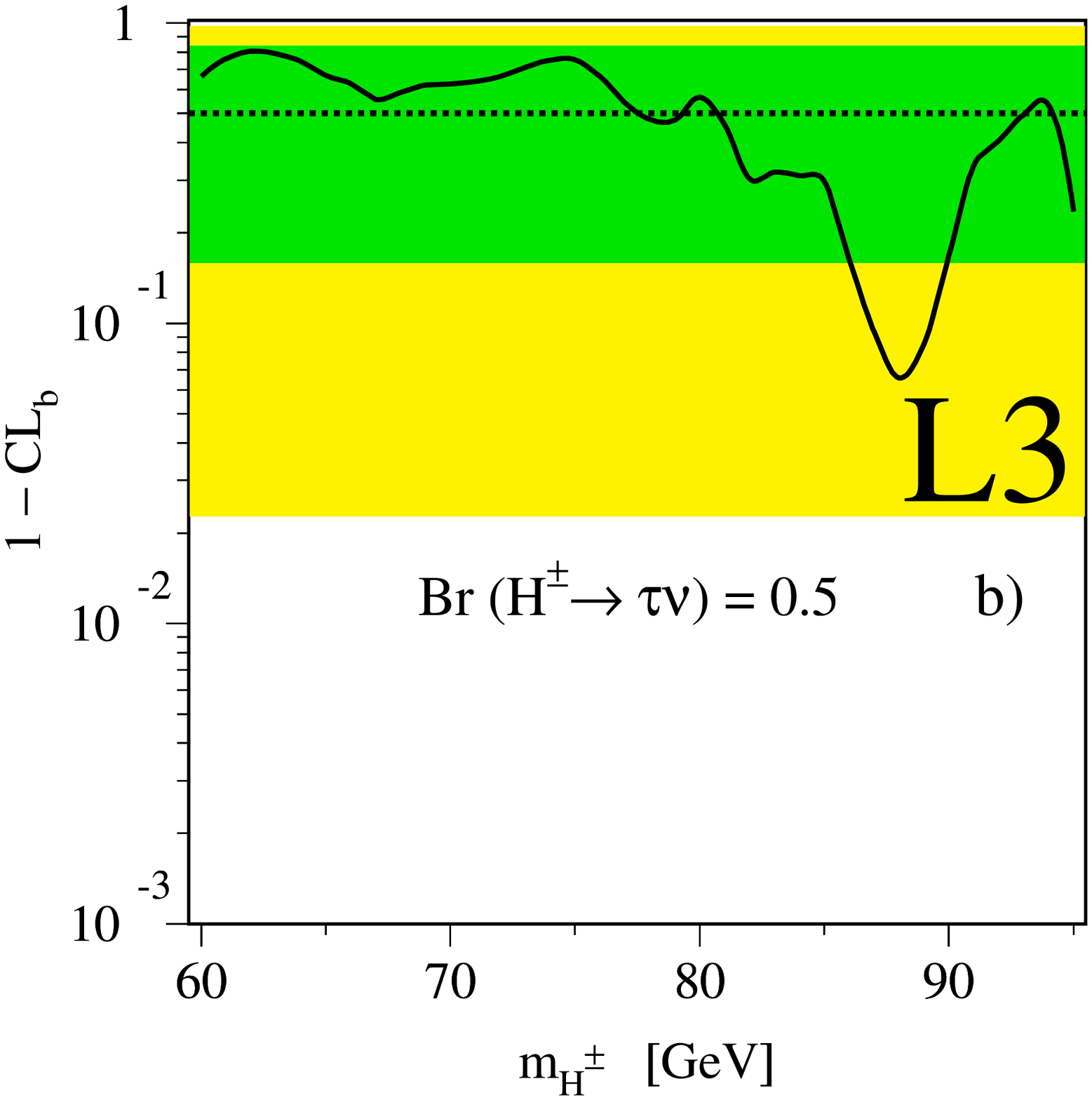,width=0.5\textwidth}}
\centerline{\epsfig{figure=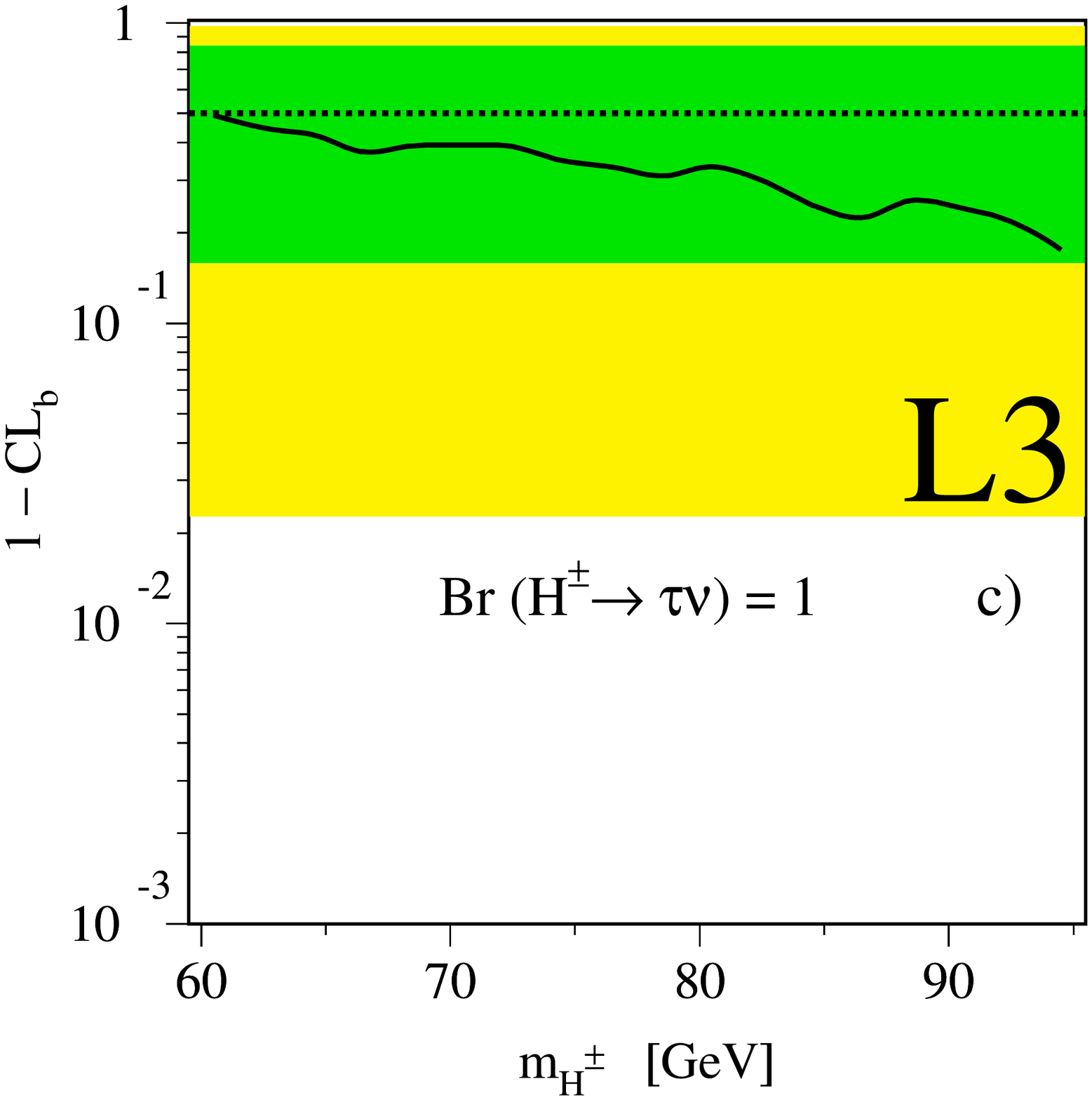,width=0.5\textwidth}}
\caption[]{\label{fig:clb} The background confidence
  level, $1-{CL_b}$, as a function of $\MHPM$ for the data (solid
  line) and for the expectation in the absence of a signal (dashed
  line), for three values of the $\Htn$ branching ratio. The shaded
  areas represent the symmetric $1 \sigma$ and $2 \sigma$ probability
  bands expected in the absence of a signal.}
\end{figure}

\begin{figure}[hp]
\centerline{\epsfig{figure=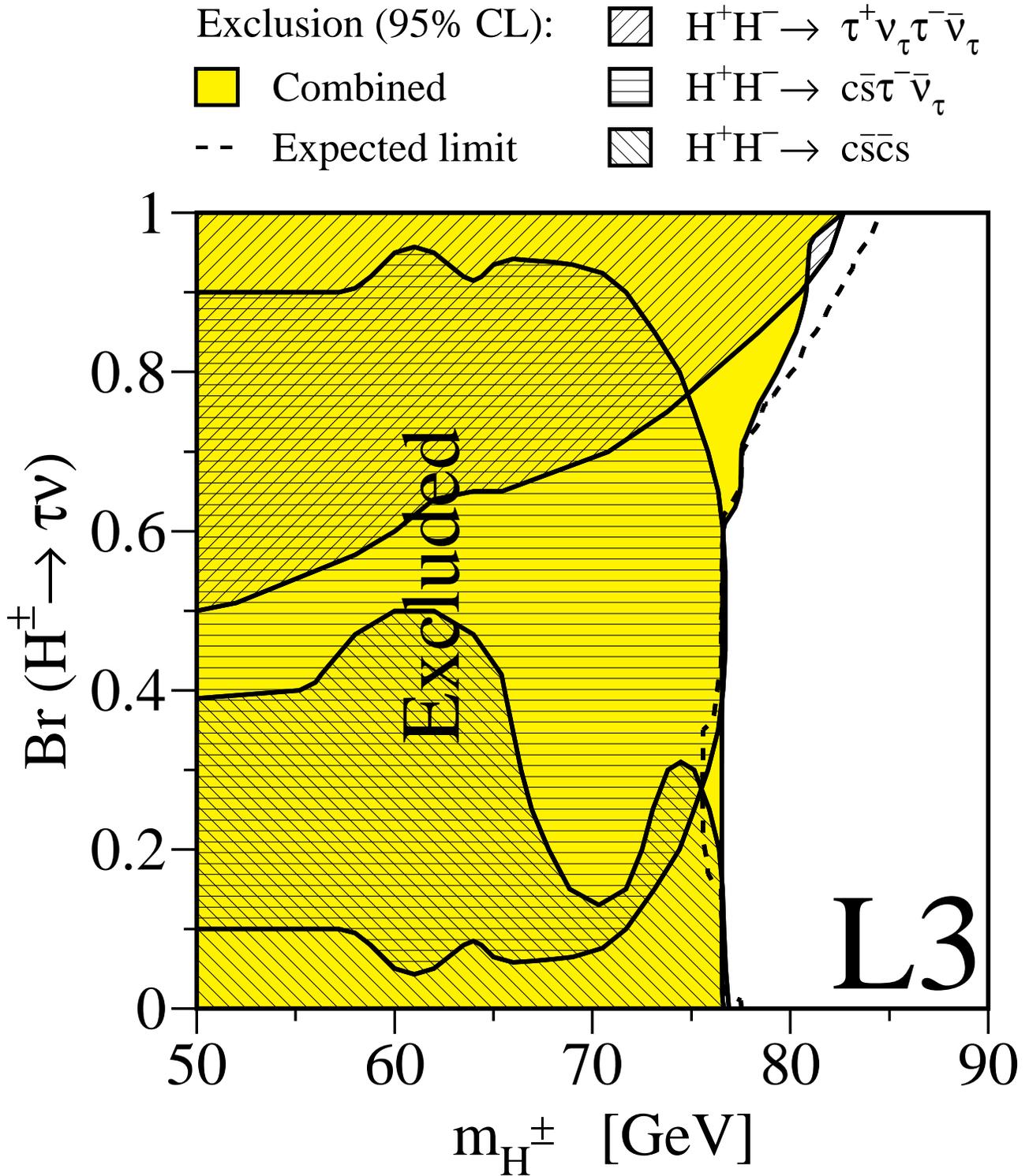,width=1.0\textwidth}}
\caption[]{\label{exclusion} Excluded regions for the charged Higgs
  boson in the plane of the $\Htn$ branching fraction versus mass, for
  the analyses of each final state and their combination. The dashed
  line indicates the median expected limit in the absence of a signal.
  Regions below $\MHPM=50\,\GeV$ are excluded by 
  data collected at the Z resonance~\cite{l3_48_50} and at $\sqrt{s}=130-183\GeV$~\cite{chhiggs_130_183}.}
\end{figure}

\end{document}

%% file: namelist274.tex
\typeout{   }     
\typeout{Using author list for paper 274 - 277 }
\typeout{$Modified: Jul 15 2001 by smele $}
\typeout{!!!!  This should only be used with document option a4p!!!!}
\typeout{   }
%
%
%
%
%
%

\newcount\tutecount  \tutecount=0
\def\tutenum#1{\global\advance\tutecount by 1 \xdef#1{\the\tutecount}}
\def\tute#1{$^{#1}$}
\tutenum\aachen            
\tutenum\nikhef            
\tutenum\mich              
\tutenum\lapp              
\tutenum\basel             
\tutenum\lsu               
\tutenum\beijing           
\tutenum\bologna           
\tutenum\tata              
\tutenum\ne                
\tutenum\bucharest         
\tutenum\budapest          
\tutenum\mit               
\tutenum\panjab            
\tutenum\debrecen          
\tutenum\dublin            
\tutenum\florence          
\tutenum\cern              
\tutenum\wl                
\tutenum\geneva            
\tutenum\hefei             
\tutenum\lausanne          
\tutenum\lyon              
\tutenum\madrid            
\tutenum\florida           
\tutenum\milan             
\tutenum\moscow            
\tutenum\naples            
\tutenum\cyprus            
\tutenum\nymegen           
\tutenum\caltech           
\tutenum\perugia           
\tutenum\peters            
\tutenum\cmu               
\tutenum\potenza           
\tutenum\prince            
\tutenum\riverside         
\tutenum\rome              
\tutenum\salerno           
\tutenum\ucsd              
\tutenum\sofia             
\tutenum\korea             
\tutenum\purdue            
\tutenum\psinst            
\tutenum\zeuthen           
\tutenum\eth               
\tutenum\hamburg           
\tutenum\taiwan            
\tutenum\tsinghua          

{
\parskip=0pt
\noindent
{\bf The L3 Collaboration:}
\ifx\selectfont\undefined
 \baselineskip=10.8pt
 \baselineskip\baselinestretch\baselineskip
 \normalbaselineskip\baselineskip
 \ixpt
\else
 \fontsize{9}{10.8pt}\selectfont
\fi
\medskip
\tolerance=10000
\hbadness=5000
\raggedright
\hsize=162truemm\hoffset=0mm
\def\r{\rlap,}
\noindent

P.Achard\r\tute\geneva\ 
O.Adriani\r\tute{\florence}\ 
M.Aguilar-Benitez\r\tute\madrid\ 
J.Alcaraz\r\tute{\madrid}\ 
G.Alemanni\r\tute\lausanne\
J.Allaby\r\tute\cern\
A.Aloisio\r\tute\naples\ 
M.G.Alviggi\r\tute\naples\
H.Anderhub\r\tute\eth\ 
V.P.Andreev\r\tute{\lsu,\peters}\
% ...
\newpage
\begin{list}{A}{\itemsep=0pt plus 0pt minus 0pt\parsep=0pt plus 0pt minus 0pt
                \topsep=0pt plus 0pt minus 0pt}
\item[\aachen]
 III. Physikalisches Institut, RWTH, D-52056 Aachen, Germany$^{\S}$
\item[\nikhef] National Institute for High Energy Physics, NIKHEF, 
     and University of Amsterdam, NL-1009 DB Amsterdam, The Netherlands
\item[\mich] University of Michigan, Ann Arbor, MI 48109, USA
\item[\lapp] Laboratoire d'Annecy-le-Vieux de Physique des Particules, 
     LAPP,IN2P3-CNRS, BP 110, F-74941 Annecy-le-Vieux CEDEX, France
\item[\basel] Institute of Physics, University of Basel, CH-4056 Basel,
     Switzerland
\item[\lsu] Louisiana State University, Baton Rouge, LA 70803, USA
\item[\beijing] Institute of High Energy Physics, IHEP, 
  100039 Beijing, China$^{\triangle}$ 
\item[\bologna] University of Bologna and INFN-Sezione di Bologna, 
     I-40126 Bologna, Italy
\item[\tata] Tata Institute of Fundamental Research, Mumbai (Bombay) 400 005, India
\item[\ne] Northeastern University, Boston, MA 02115, USA
\item[\bucharest] Institute of Atomic Physics and University of Bucharest,
     R-76900 Bucharest, Romania
\item[\budapest] Central Research Institute for Physics of the 
     Hungarian Academy of Sciences, H-1525 Budapest 114, Hungary$^{\ddag}$
\item[\mit] Massachusetts Institute of Technology, Cambridge, MA 02139, USA
\item[\panjab] Panjab University, Chandigarh 160 014, India.
\item[\debrecen] KLTE-ATOMKI, H-4010 Debrecen, Hungary$^\P$
\item[\dublin] Department of Experimental Physics,
  University College Dublin, Belfield, Dublin 4, Ireland
\item[\florence] INFN Sezione di Firenze and University of Florence, 
     I-50125 Florence, Italy
\item[\cern] European Laboratory for Particle Physics, CERN, 
     CH-1211 Geneva 23, Switzerland
\item[\wl] World Laboratory, FBLJA  Project, CH-1211 Geneva 23, Switzerland
\item[\geneva] University of Geneva, CH-1211 Geneva 4, Switzerland
\item[\hefei] Chinese University of Science and Technology, USTC,
      Hefei, Anhui 230 029, China$^{\triangle}$
\item[\lausanne] University of Lausanne, CH-1015 Lausanne, Switzerland
\item[\lyon] Institut de Physique Nucl\'eaire de Lyon, 
     IN2P3-CNRS,Universit\'e Claude Bernard, 
     F-69622 Villeurbanne, France
\item[\madrid] Centro de Investigaciones Energ{\'e}ticas, 
     Medioambientales y Tecnol\'ogicas, CIEMAT, E-28040 Madrid,
     Spain${\flat}$ 
\item[\florida] Florida Institute of Technology, Melbourne, FL 32901, USA
\item[\milan] INFN-Sezione di Milano, I-20133 Milan, Italy
\item[\moscow] Institute of Theoretical and Experimental Physics, ITEP, 
     Moscow, Russia
\item[\naples] INFN-Sezione di Napoli and University of Naples, 
     I-80125 Naples, Italy
\item[\cyprus] Department of Physics, University of Cyprus,
     Nicosia, Cyprus
\item[\nymegen] University of Nijmegen and NIKHEF, 
     NL-6525 ED Nijmegen, The Netherlands
\item[\caltech] California Institute of Technology, Pasadena, CA 91125, USA
\item[\perugia] INFN-Sezione di Perugia and Universit\`a Degli 
     Studi di Perugia, I-06100 Perugia, Italy   
\item[\peters] Nuclear Physics Institute, St. Petersburg, Russia
\item[\cmu] Carnegie Mellon University, Pittsburgh, PA 15213, USA
\item[\potenza] INFN-Sezione di Napoli and University of Potenza, 
     I-85100 Potenza, Italy
\item[\prince] Princeton University, Princeton, NJ 08544, USA
\item[\riverside] University of Californa, Riverside, CA 92521, USA
\item[\rome] INFN-Sezione di Roma and University of Rome, ``La Sapienza",
     I-00185 Rome, Italy
\item[\salerno] University and INFN, Salerno, I-84100 Salerno, Italy
\item[\ucsd] University of California, San Diego, CA 92093, USA
\item[\sofia] Bulgarian Academy of Sciences, Central Lab.~of 
     Mechatronics and Instrumentation, BU-1113 Sofia, Bulgaria
\item[\korea]  The Center for High Energy Physics, 
     Kyungpook National University, 702-701 Taegu, Republic of Korea
\item[\purdue] Purdue University, West Lafayette, IN 47907, USA
\item[\psinst] Paul Scherrer Institut, PSI, CH-5232 Villigen, Switzerland
\item[\zeuthen] DESY, D-15738 Zeuthen, Germany
\item[\eth] Eidgen\"ossische Technische Hochschule, ETH Z\"urich,
     CH-8093 Z\"urich, Switzerland
\item[\hamburg] University of Hamburg, D-22761 Hamburg, Germany
\item[\taiwan] National Central University, Chung-Li, Taiwan, China
\item[\tsinghua] Department of Physics, National Tsing Hua University,
      Taiwan, China
\item[\S]  Supported by the German Bundesministerium 
        f\"ur Bildung, Wissenschaft, Forschung und Technologie
\item[\ddag] Supported by the Hungarian OTKA fund under contract
numbers T019181, F023259 and T037350.
\item[\P] Also supported by the Hungarian OTKA fund under contract
  number T026178.
\item[$\flat$] Supported also by the Comisi\'on Interministerial de Ciencia y 
        Tecnolog{\'\i}a.
\item[$\sharp$] Also supported by CONICET and Universidad Nacional de La Plata,
        CC 67, 1900 La Plata, Argentina.
\item[$\triangle$] Supported by the National Natural Science
  Foundation of China.
\end{list}
}
\vfill
